# Electric field control of magnetic phase transitions in $Ni_3V_2O_8$

P. Kharel[1], C. Sudakar[1], A.B. Harris[2], R. Naik[1], G. Lawes[1]

[1]Department of Physics and Astronomy, Wayne State University, Detroit, MI 48201
[2]Department of Physics, University of Pennsylvania, Philadelphia, PA 19104

*Abstract*

We report on the electric field control of magnetic phase transition temperatures in multiferroic $Ni_3V_2O_8$ thin films. Using magnetization measurements, we find that the phase transition temperature to the canted antiferromagnetic state is suppressed by 0.2 K in an electric field of 30 MV/m, as compared to the unbiased sample. Dielectric measurements show that the transition temperature into the magnetic state associated with ferroelectric order increases by 0.2 K when the sample is biased at 25 MV/m. This electric field control of the magnetic transitions can be qualitatively understood using a mean field model incorporating a tri-linear coupling between the magnetic order parameters and spontaneous polarization.

Ferroelectric order and magnetic order are typically promoted by mutually exclusive materials properties. However, in certain special classes of materials long range magnetic and ferroelectric order can develop concomitantly, with substantial coupling between the two [1]. It has been established that in a number of these magnetoelectric multiferroics the ferroelectric polarization can be controlled by an external magnetic field. An external magnetic field produces an orthogonal rotation of polarization in $TbMnO_3$ [2, 3], a polarization reversal in $TbMn_2O_5$ [4] and a complete suppression of polarization in $Ni_3V_2O_8$ [5, 6]. Several models have been proposed for systems that develop magnetic and ferroelectric order simultaneously at a single phase transition, including Landau symmetry analysis for $Ni_3V_2O_8$ and other similar systems [5, 7-9], a theory of ferroelectricity in inhomogeneous magnets based on Ginzburg-Landau symmetry analysis [10] and a proposal of possible microscopic mechanisms for ferroelectricity in nonlinear magnets [11-13].

There is also considerable interest in understanding how an electric field can be used to control a magnetic structure, in part because of the possibility of developing novel devices based on voltage switchable magnetic memory and highly tunable dielectric constants [14]. While the electric field control of transition temperature in ferromagnetic semiconductors [15] and electrostatic modulation of correlated electron behavior in complex oxides [16] have been widely investigated, there are fewer studies on electric field control involving magnetoelectric multiferroics. It has been demonstrated that the magnetic structure can be modified electrostatically in certain multiferroic materials as demonstrated by optical studies on $HoMnO_3$ [17] photoemission electron microscopy in $BiFeO_3$ [18] and neutron scattering measurements in $TbMnO_3$ [19]. Very recent studies have shown that the net magnetization can be switched using an applied electric field in $La_{0.67}Sr_{0.33}MnO_3/BaTiO_3$ epitaxial heterostructures [20]. In the following, we present studies establishing that due to strong spin-charge coupling, an applied electric field can shift the magnetic transition temperatures in $Ni_3V_2O_8$.

$Ni_3V_2O_8$ (NVO) is a widely studied multiferroic material [5, 21-25]. It is a magnetic insulator consisting planes of spin-1 $Ni^{2+}$ ions arranged in a kagome staircase lattice. NVO undergoes a series of magnetic phase transitions at zero magnetic field. At $T_H$=9.1 K it undergoes a magnetic transition into a high temperature incommensurate (HTI) phase, in which the spins are mainly collinearly ordered with sinusoidally varying amplitude as described by the complex-valued order parameter $\sigma_{HTI}$. At $T_L$=6.3 K NVO undergoes a second phase transition into the low temperature incommensurate (LTI) phase in which the spins develop additional transverse order governed by a second complex order parameter $\sigma_{LTI}$. At lower temperatures, NVO undergoes two additional phase transitions into slightly different canted antiferromagnetic (CAF) phases at $T_C$=3.9 K and $T_{C'}$=2.2 K [21]. Most significantly for this work, NVO also exhibits ferroelectric order in the LTI phase with a spontaneous polarization along the *b*-axis. NVO is highly insulating and thin films can be easily prepared on a variety of substrates [26], making it an ideal material for studies on electric field control of magnetic phase transitions in multiferroic materials. Theoretical predictions suggest that the ferroelectric LTI phase in NVO should be stabilized in an electric field [12].

We have prepared thin films of NVO by RF sputter deposition using a dense polycrystalline target prepared by standard solid state techniques. Films with thickness from 500 nm to 600 nm were deposited on *a*-cut As doped conducting silicon at room temperature. These films were annealed at $1000^0$C for two hours in air to improve the crystallinity. More detailed discussion of thin film NVO preparation is presented elsewhere [26].

We investigated crystal structure of the films using powder X-Ray Diffraction (XRD) on a Rigaku RU2000 diffractometer. XRD pattern were collected at room temperature with copper $K_\alpha$ radiation in θ-2θ mode. The XRD pattern of the as-prepared films (not shown) showed only a diffused background, consistent with an amorphous sample. The diffraction peaks for the annealed samples, shown in the upper panel of Figure 1, can be completely indexed to the NVO structure (ref. PDF card no 70-1394), with no impurity phases observed. Although the films are polycrystalline, the presence of only the (020), (040) and (080) peaks suggests that they are highly *b*-axis oriented. All three peaks are slightly shifted towards higher 2θ values, consistent with tensile strain on the film. Raman spectra collected for the thin film (not shown) exactly match up with bulk $Ni_3V_2O_8$. We imaged the films using Scanning Electron Microscopy (SEM), which shows that there are a number of large faceted grains of approximately 2μm x 1μm on a uniform NVO background, as shown in the lower panel of Figure 1. Cross sectional SEM reveals that this background consists of a layer of closely packed rods. We estimate the thickness of the samples to be between 500 nm to 600 nm, using the cross sectional SEM micrograph shown in the lower panel of Figure 1.

We characterized the magnetic properties of the thin films using a Quantum Design SQUID magnetometer. Figure 2 shows the temperature dependence of the magnetization measured in a field of H=500 Oe on both cooling and warming, as indicated by the arrows. The closed symbols show the magnetization with no bias voltage applied, while the open symbols plot the magnetization with a bias field of 30 MV/m applied perpendicular to the film thickness, for which we expect to be predominantly along the *b*-axis for the grains. The magnetic anomaly is consistent with the first order phase transition from the LTI to CAF phase observed in bulk $Ni_3V_2O_8$ samples [25]. This hysteresis is much broader than for bulk samples, with a width of approximately 0.5 K as compared to only 0.02 K for bulk samples. This may be consistent with the thin film samples being comprised of disordered grains. Using a film thickness 500 nm having an area 0.3 $cm^2$, the change in magnetic moment at the transition corresponds to approximately 0.001 $\mu_B$/Ni, which is a factor of three smaller than the value reported for bulk NVO [26]. There is an additional very small magnetic anomaly at approximately T=6.5 K. We believe that this feature is not intrinsic to the $Ni_3V_2O_8$ sample, as a magnetic feature appears at this temperature in the magnetization of the bare silicon substrate. We plot an M(T) curve for the bare silicon substrate in the inset panel to Figure 2, which clearly shows a sharp drop in moment at a temperature consistent with the very small anomaly in the $Ni_3V_2O_8$ thin film samples.

The most significant feature observed in the M(T) curves in Figure 2 is the bias voltage dependence of the magnetization. Applying an electric field perpendicular to the film thickness (approximately parallel to the *b*-axis) suppresses the LTI to CAF phase transition temperature. This suppression of the magnetic transition temperature reaches approximately 0.2 K in an electric field of 30 MV/m. We believe that this effect is intrinsic, rather than arising from incidental sample heating, for several reasons. Above the LTI transition temperature, there is no bias dependence of the magnetization. Furthermore, although the silicon substrate is expected to be in excellent thermal contact with the NVO film, there is no shift in the temperature of the substrate magnetic anomaly. Finally, as we will show in the following, there is no evidence for bias voltage induced heating in dielectric measurements on the $Ni_3V_2O_8$ thin films. This voltage induced shift in the magnetization curves is strong evidence that LTI spin structure in NVO can be stabilized by an applied electric field. We note that since the CAF spin structure does not transform under a subgroup of the LTI magnetic symmetry, this transition remains a real thermodynamic phase transition under the application of an electric field. This is distinct from the HTI to LTI boundary, where applying an electric field breaks a symmetry in the HTI phase leading to a cross-over transition.

To further probe the electric field control of the LTI magnetic phase, we conducted extensive dielectric measurements at the LTI to HTI phase boundary while applying external electric and magnetic fields. There is no significant magnetic anomaly associated with the HTI to LTI magnetic phase boundary, but the dielectric constant is sharply peaked at this transition [5]. Because the LTI phase is multiferroic, we can determine the *magnetic* transition temperature by tracking the *dielectric* anomaly. We plot the NVO dielectric constant as a function of temperature under different representative external magnetic and electric fields in Figure 3. For these measurements, both the magnetic field and electric field are applied in the perpendicular direction, along the *b*-axis for the majority of the NVO grains comprising the film. With no external fields applied, the dielectric constant shows a sharp peak at $T_L$=6.3 K. This peak is significantly broader than what is measured in bulk samples. This may be attributed in part to the finite grain size in our sample, but we also expect significant broadening to occur due to symmetry breaking induced by the external electric field, similar to what is observed in magnetic susceptibilities at a ferromagnetic phase transition in a finite magnetic field. When an electric field E=25 MV/m is applied to the sample, this peak shifts up in temperature by approximately $\Delta T_L$= 0.2 K, while becoming smaller and broader. In order to confirm that this single peak indicates the onset of both the LTI magnetic structure and ferroelectric order, we also measured the NVO dielectric constant under the simultaneous application of an electric field *E*=25 MV/m and magnetic field $\mu_0$H=7.5 T. The single peak indicating the development of the LTI magnetic structure shifts down in temperature, confirming that this single transition is sensitive to both magnetic and electric fields. We note that these shifts cannot be caused simply by sample heating, since this would result only in negative temperature shifts, as measured by the system thermometer.

We summarize the results of our measurements on the electric field dependence of the LTI to CAF magnetic phase transition and the electric and magnetic field dependence of the HTI to LTI magnetic phase transition in Figures 4a and 4b respectively. The dashed lines are guides to the eye and are discussed in the following. Figure 4a plots the bias voltage versus approximate LTI to CAF phase transition temperature, as estimated by the E-field induced shift in the M(T) curves measured at H=500 Oe, as illustrated in Figure 2. Because this transition is highly hysteretic, it is difficult to accurately determine the transition temperatures from these curves, although there is considerably less error in determining the relative change in transition temperature. The solid circles in Figure 4b plots the bias voltage versus the HTI to LTI phase transition (*E*=0) or cross-over (finite *E*) temperature at zero magnetic field, determined by dielectric measurements similar to those plotted in Figure 3. In both cases, applying an electric field parallel to the *b*-axis in NVO stabilizes the ferroelectric LTI magnetic structure. We also confirm that an external magnetic field shifts the HTI to LTI transition temperature, as shown by the solid triangles in Figure 4b. Applying a magnetic field along the b direction suppresses the magnetic transition temperature, which is increased when a bias voltage is applied on top of the external magnetic field (open star in Figure 4b).

These results can be understood qualitatively within the framework of a mean field trilinear coupling between the LTI and HTI magnetic order parameters and the spontaneous polarization. Estimates from this model for the electric field induced shift in the LTI to CAF phase transition temperature, based on magnetic and thermodynamic measurements on bulk NVO, predict that the transition temperature should decrease linearly with applied E-field, with a slope of $1.5 \times 10^{-7}$ Km/V. The dashed line in Figure 4a is intended as a guide to the eye to test this model; the slope of this line is approximately $0.05 \times 10^{-7}$ Km/V. While this shift is significantly smaller than what is predicted from this simple model, the qualitative behavior is similar. Furthermore, the $Ni_3V_2O_8$ films in this investigation are rough, so it is likely that the electric field at each grain is not completely aligned along the b-axis, which could reduce the shift in temperature. Based on these results, we suggest that improving the quality of the thin film samples could potentially increase this electric field induced shift in the magnetic transition by an order of magnitude.

The phase boundary at the HTI to LTI transition plotted in Figure 4b suggests that the shift in transition temperature is not linear with the applied electric field. The dashed line plots a dependence $E \sim (\Delta T)^2$, which is closer to the measured phase boundary. In the framework of this mean field model, we suppose that the applied electric field along the b-axis produces an electric polarization ($P_b$), which in turn couples to the LTI phase magnetic order parameter ($\sigma_{LTI}$).

The free energy near the LTI-HTI phase boundary can be expressed as:

$$F = \tfrac{1}{2}\chi_E^{-1}P^2 + a\left(\langle\sigma_{HTI}^*\rangle\sigma_{LTI} + \langle\sigma_{HTI}\rangle\sigma_{LTI}^*\right)P_b + \tfrac{1}{2}(T - T_L)|\sigma_{LTI}|^2 - \vec{E}\cdot\vec{P} \quad (1)$$

where $\chi_E$ is the electric susceptibility, and a is a constant. Since the LTI-HTI phase boundary is far from the transition to the HTI phase, we replace the HTI magnetic order parameter, $\sigma_{HTI}$, by its average value $\langle\sigma_{HTI}\rangle$. This expression is discussed in detail in Ref. [9]. The true LTI order parameter, $\tilde{\sigma}$, is found by diagonalizing the quadratic form in $P_b$ and $\sigma_{LTI}$, and is therefore given by:

$$\tilde{\sigma} = b\sigma_{LTI} + b^*\sigma_{LTI}^* + cP_b \quad (2)$$

From this expression, one sees that an applied electric field acts as a field conjugate to the real order parameter $\tilde{\sigma}$. Because the coefficient a in Eq. (1) is small, as evidenced by the very small spontaneous polarization in NVO, the contribution of P to the real order parameter is also small. Near the LTI-HTI phase boundary the order parameter can be written as:

$$\langle\tilde{\sigma}\rangle = t^\beta F\left(Et^{-\beta-\gamma}\right) \quad (3)$$

where $\beta$ and $\gamma$ are the usual critical exponents for the magnetization and susceptibility respectively [27]. This implies that $E\sim(\Delta T)^{\beta+\gamma}$. Within mean field theory $\beta=1/2$ and $\gamma=1$, which would lead to $E\sim(\Delta T)^{1.5}$, but fluctuations lead to a significant increase in $\gamma$ and usually a smaller decrease in $\beta$, so that typically one expects $\gamma+\beta\sim1.7$. Using the scaling equation $\beta+\gamma=2-\alpha-\beta$, with $\alpha$ the specific heat exponent (which need not be positive), leads to a similar conclusion. From these arguments, it is not surprising that the E-T phase diagram illustrated in Figure 4b suggests that $\beta+\gamma$ is close to 2. Because the real order parameter, $\tilde{\sigma}$, couples to an applied electric field, we believe that the symmetry of the HTI phase is broken by the external electric field, so that the HTI to LTI transition in finite E-fields is a cross-over effect rather than a real thermodynamic phase transition. However, since the magnitude of the polarization coefficient in Eq. (2) is small, an external electric field provides only a very small field conjugate to the real order parameter, so the actual HTI to LTI phase transition is likely to be only slightly smeared under an applied bias voltage.

In conclusion, we have demonstrated that the transition temperature for magnetic phases in $Ni_3V_2O_8$ thin films can be controlled using an external electric field. Because of the coupling between the magnetic and ferroelectric order parameters, the LTI phase is stabilized by the application of an electric field, as confirmed by magnetic and dielectric measurements. This effect is complementary to the magnetic field control of the ferroelectric polarization observed in many other multiferroic systems. While the E-field induced shifts in the magnetic transition temperatures were rather small, we expect that higher quality samples having better orientation may show stronger coupling. The ability to switch the magnetization using a bias voltage is an important characteristic for many of the proposed potential applications for multiferroic materials.

This work was supported by the NSF under DMR-0644823, and by the Wayne State University Institute for Manufacturing Research.

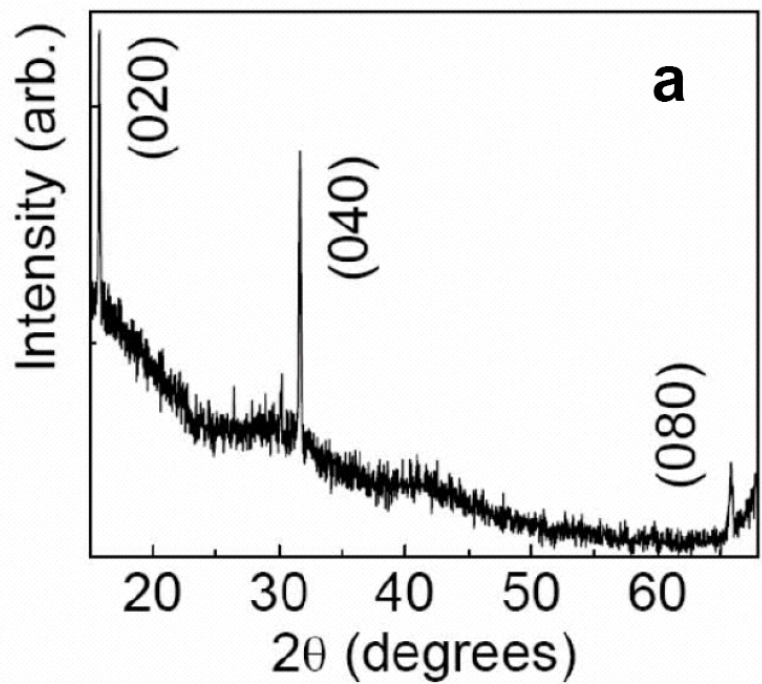
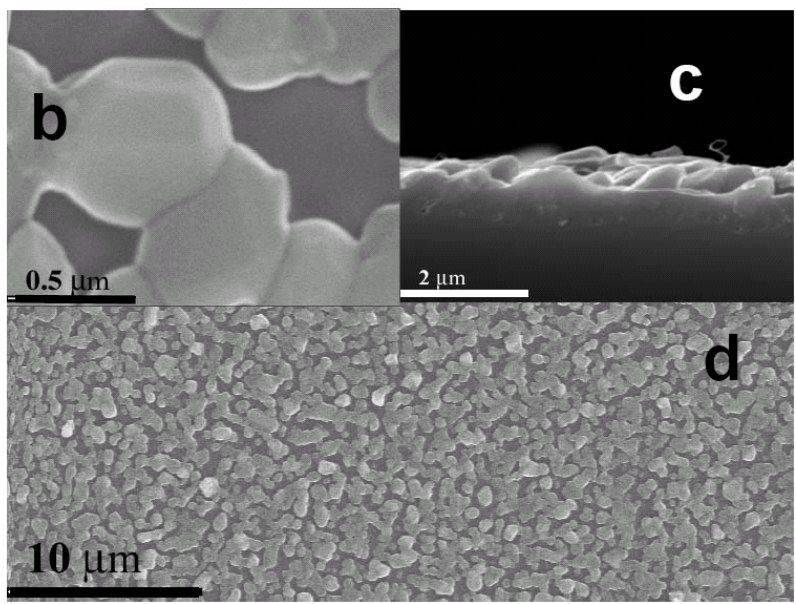

**Figure 1**. (a) XRD pattern for Ni$_3$V$_2$O$_8$ thin film. (b) SEM image (scale bar is 500 nm), c) Cross-section SEM image (scale bar is 2 µm), d) SEM image (scale bar is 10 µm).

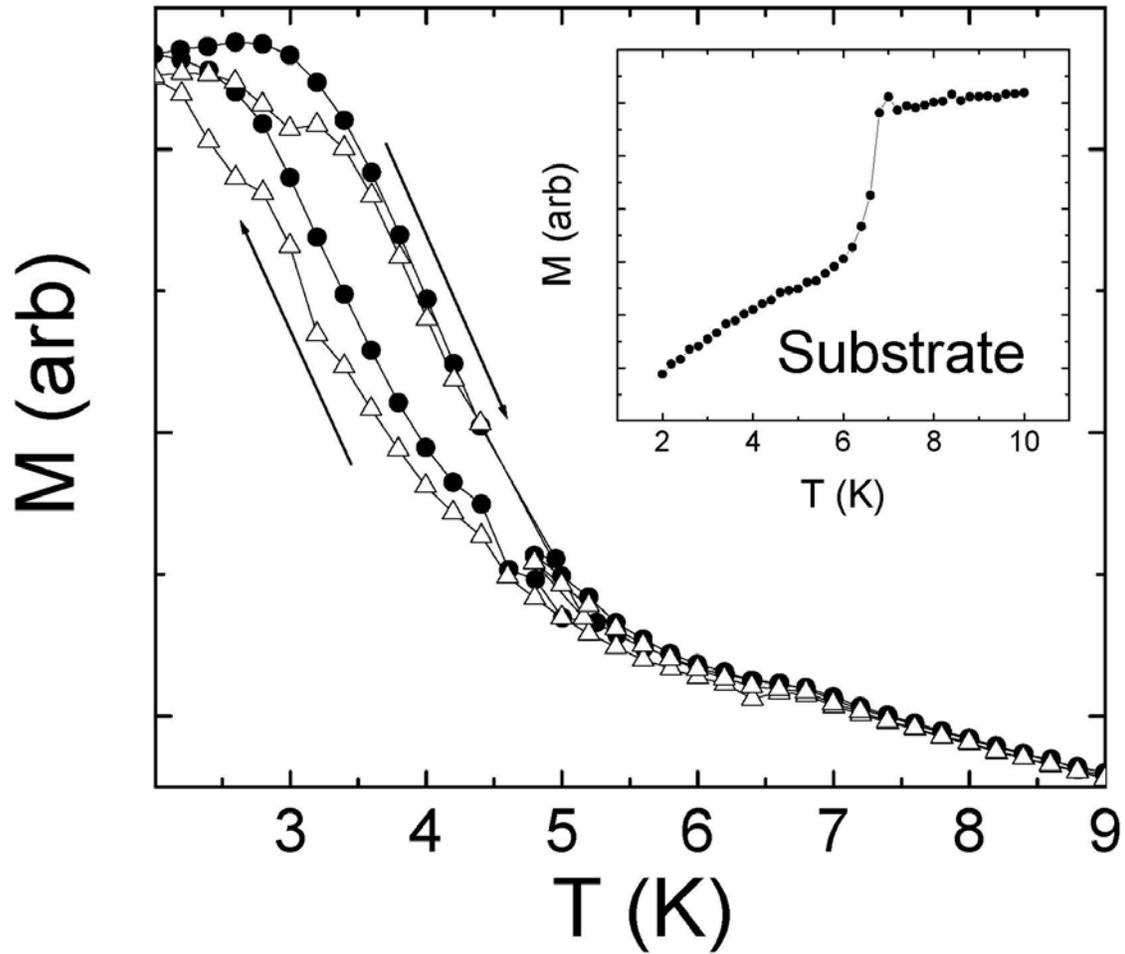

**Figure 2**. M(T) curves for $Ni_3V_2O_8$ at E=0 (closed symbols) and E=30 MV/m (open symbols) with H=500 Oe. The arrows indicated heating and cooling. INSET: M(T) for the silicon substrate measured at H=500 Oe.

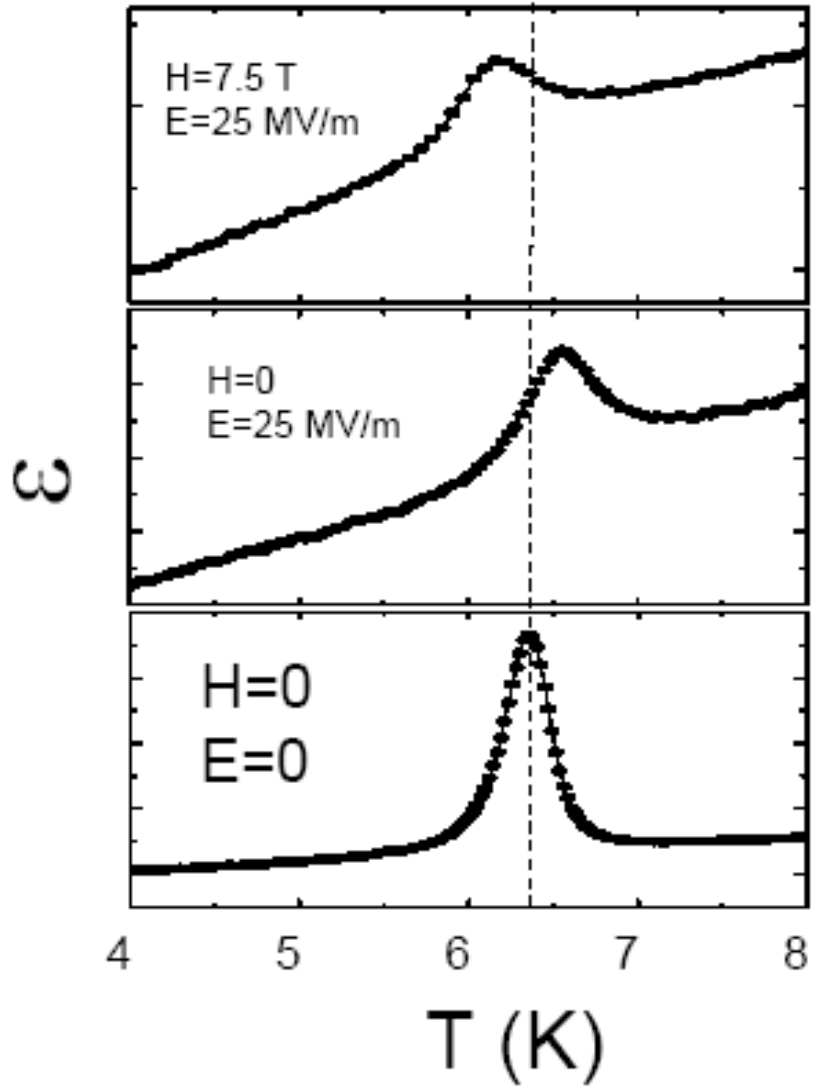

**Figure 3**. Dielectric constant of $Ni_3V_2O_8$ thin film measured at different electric and magnetic fields. (a) $\mu_0H$=7.5 T, E=25 MV/m, (b) H=0, E=25 MV/m, and (c) H=E=0.

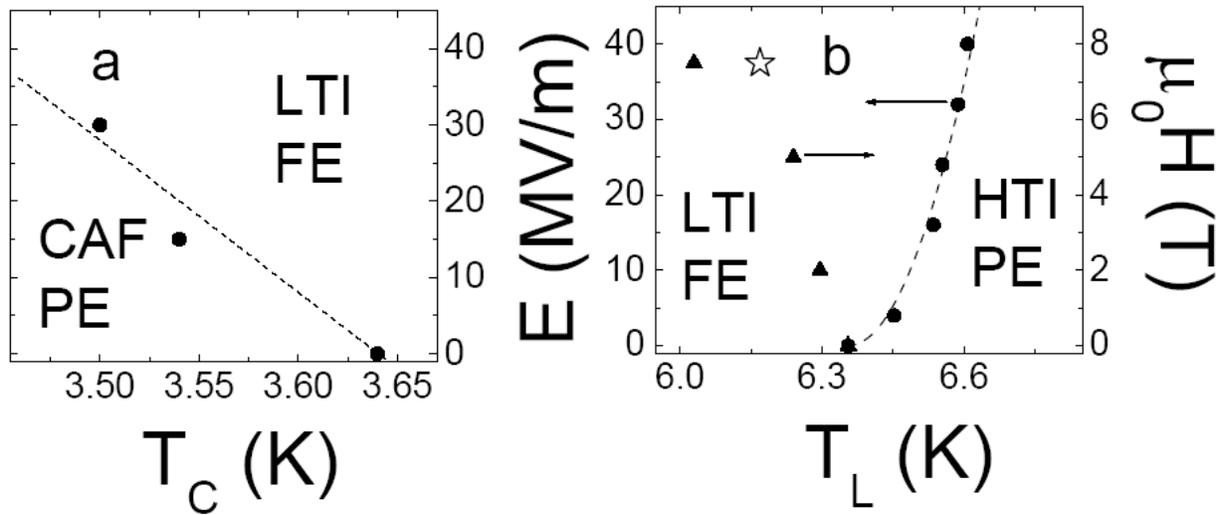

**Figure 4**. (a) E-T phase diagram for $Ni_3V_2O_8$ at the LTI to CAF phase transition, measured at H=0. (b) E-H-T phase diagram near the HTI to LTI phase transition. The solid triangles show $T_L$ measured as a function of magnetic field in zero electric field, while the solid circles show $T_L$ measured as a function of electric field at zero magnetic field. The open star shows the transition temperature measured with both a magnetic and electric field applied. The dashed lines are guides to the eye discussed in the text.